# 2D magnetoelectric multiferroics in MnSTe/In$_2$Se$_3$ heterobilayer with ferroelectrically controllable skyrmions

Kaiying Dou, Wenhui Du, Ying Dai,* Baibiao Huang and Yandong Ma*

School of Physics, State Key Laboratory of Crystal Materials, Shandong University, Shandanan Street 27, Jinan 250100, China

*Email: daiy60@sina.com (Y.D.); yandong.ma@sdu.edu.cn (Y.M.)

**Abstract:**

The magnetoelectric effect and skyrmions are two fundamental phenomena in the field of condensed-matter physics. Here, using first-principles calculations and Monte-Carlo simulations, we propose that strong magnetoelectric coupling can be demonstrated in a multiferroic heterobilayer consisting of two-dimensional (2D) MnSTe and α-In$_2$Se$_3$. As the electric polarization in ferroelectric α-In$_2$Se$_3$ is switched, the creation and annihilation of topological magnetic phase can be achieved in this multiferroic heterobilayer, giving rise to the intriguing ferroelectrically controllable skyrmions. This feature is further revealed to be closely related to the physical quantity of $D^2/|KJ|$, which is generally applicable for describing the required conditions of such physics. Moreover, the evaluations of their topological magnetic phases with temperature are systematically discussed. These insights not only greatly enrich the research on 2D magnetoelectric multiferroics, but also pave a promising avenue to realize new skyrmionic device concepts.

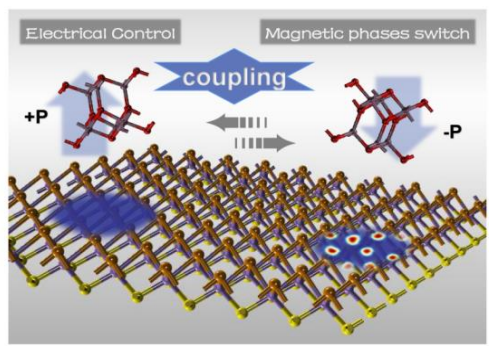

## Introduction

Two-dimensional (2D) multiferroics are a special class of materials that hold simultaneously two or more primary ferroic (i.e., ferromagnetic, ferroelectric and ferroelastic) orders[1-6]. Among numerous multiferroics, magnetoelectric multiferroics has attracted intensive attention because of its novel physics and potential applications in next-generation of spintronics and memory devices[5-12]. The current research on magnetoelectric multiferroics in 2D lattice has been mainly established in the paradigm of ferroelectrically manipulation of magnetic ground state[9,13-15], magnetization orientation[11,16-17] or magnetic moment distribution[5,18-19], leaving other magnetic behaviors less explored in this field[20-22]. Ideal 2D magnetoelectric multiferroics requires ferromagnetic and ferroelectric orders to be strongly coupled[23]. Until now, however, reported 2D magnetoelectric multiferroics are rather scarce[11-19] and most with coexisting yet loosely coupled ferromagnetic and ferroelectric orders[12-19] owning to the inherent exclusion between ferroelectricity and magnetism. 2D multiferroics with strong magnetoelectric coupling, especially on novel magnetic properties, remains highly desired from both scientific and technological impact.

Compared with conventional magnetic properties, magnetic skyrmions are a newly emerging topological magnetic order, which, however, spur rapid development in both fundamental research and device applications[24-27]. The unique fingerprint of magnetic skyrmions can be characterized by the spin swirling vortex-like texture, which is protected by a real-space topological number[28-29]. Magnetic skyrmions are first proposed experimentally in B20 compounds, such as MnSi[27] and FeGe[30]. Recently, with the advance on long-range magnetism in 2D CrI$_3$[31], Cr$_2$Ge$_2$Te$_6$[32] and VSe$_2$[33-34], impressive progress has been made in realizing magnetic skyrmions in 2D lattice. Typical examples include 2D Janus TMDs[35-37], CrGe(Se, Te)$_3$[38], Cr(I,X)$_3$(X=Cl, Br)[39], CrN[40] and so on[41-43]. To make skyrmionic devices, the essential step is to control skyrmions[44-47]. With respect to the dynamic strategies[48], undoubtedly, the possible coupling between ferroelectrics and magnetic skyrmions, which adds skyrmions into magnetoelectric effect, could provide an unprecedent opportunity to control skyrmions in a robust way and enable novel physical phenomena. Through highly valuable, so far, 2D magnetoelectric multiferroics with ferroelectrically controllable skyrmions is rarely explored[20-21].

In this work, we design a multiferroic heterobilayer consisting of magnetic MnSTe and ferroelectric α-In$_2$Se$_3$. Using first-principles calculations and Monte Carlo simulations, we find that significant magnetoelectric effect can be realized in this multiferroic heterobilayer. Upon reversing the electric polarization in ferroelectric α-$In_2Se_3$, the topological magnetic phase can be switched on and off, which leads to the exotic ferroelectrically controlled creation and annihilation of magnetic skyrmions. We further unveil that such phenomenon closely correlates to the physical quantity of $D^2/|KJ|$, and this quantity can be employed to estimate the required conditions of such physics. Finally, we discuss the effect of temperature on the skyrmions number and topological magnetic phases of such multiferroic heterobilayer. The explored phenomena and mechanism are useful for fundamental research in 2D magnetoelectric multiferroics and skyrmionics.

**Computational Method**

Our first-principles calculations are performed based on density functional theory as implemented in Vienna ab initio simulation package[49]. The projector augmented wave method[50] is adopted to treat the ionic potential. The Perdew-Burke-Ernzerhof functional of generalized gradient approximation[51] is used for the exchange-correlation interactions. To describe the strongly correlated electrons in 3d orbital of Mn atom, we employ effective Hubbard[52] U = 2 eV as reported in previous works[36]. The convergence tolerances for the residual force and energy are set to 0.001 eV/Å and $10^{-6}$ eV, respectively. The plane-wave cutoff energy is set to 520 eV. The Monkhorst-Pack k-point mesh[53] of 13 × 13 × 1 is adopted to sample the Brillion zone. The DFT-D3 method[54] is employed for treating van der Waals interaction in all calculations. The ferroelectric transition pathway and energy barrier are calculated on the basis of the nudged elastic band (NEB) method[55].

Using the magnetic parameters obtained from first-principles calculations, we employ parallel tempering Monte Carlo (MC) simulations[56] with the Metropolis algorithm to obtain the energy minimum spin textures. The spin textures are obtained based on a 150×150×1 supercell with 100000 MC steps performed at each temperature (from 660 K cooled down to the investigated low temperature).

**Results and Discussion**

The heterobilayer that we designed consists of a monolayer MnSTe and a monolayer α-$In_2Se_3$. While monolayer MnSTe harbors topological magnetic order[35-36], monolayer α-$In_2Se_3$ is known for its out-of-plane ferroelectric polarization[57]. The lattice constants of MnSTe and α-In$_2$Se$_3$ are optimized to be 3.55 Å and 4.02 Å, respectively, which agree well with the previous studies[36,57]. For the heterobilayer, the $2 \times 2$ supercell of MnSTe is used to match the $\sqrt{3} \times \sqrt{3}$ supercell of α-In$_2$Se$_3$, which results in a rather small lattice mismatch of 1.9 %. The contacted surface of MnSTe with α-In$_2$Se$_3$ could be either S- or Te-terminated surface, and they share roughly similar results. In the following, we only focus on S-terminated case. For more information on Te-terminated case, please see Supporting Information. Given the switchable polarization of α-In$_2$Se$_3$, the heterobilayer naturally displays two types: the first one with polarization pointing to the interface (P+), while the second one with polarization pointing away from the interface (P-). For each type, different stacking patterns are considered, and the energy minimum patterns are shown in **Fig. 1a**.

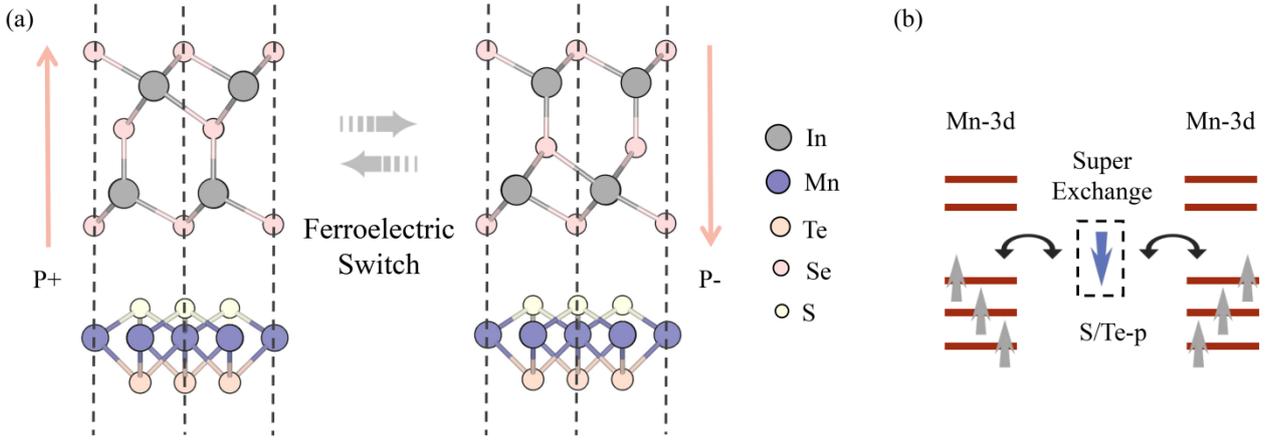

**Fig. 1** (a) Crystal structures of MnSTe/In$_2$Se$_3$ heterobilayer under P+ and P- configurations. (b) Schematic diagram of *d*-orbital splitting under distorted octahedral coordination environment and the super exchange coupling between *3d* electrons of the Mn atoms in the heterobilayer.

The interlayer distances for P+ and P- phases of the heterobilayer are found to be 0.106 and 0.108 Å, respectively, and the corresponding electric polarizations are calculated to be $3.263 \times 10^{-12}$ and $3.092 \times 10^{-11}$ C/m$^2$. Obviously, they display different absolute values for the electric polarizations. Nevertheless, as we will show later, P+ and P- phases can be considered as two ferroelectric states of the heterobilayer and the ferroelectric transition between them is feasible. The valence electron configuration of Mn atom is $3d^54s^2$. By coordinating with three S and three Te atoms, four electrons are transferred to the neighboring atoms, leading to the electron configuration of $3d^34s^0$. Under the distorted octahedral coordination environment, the *d* orbitals roughly split into two groups, i.e., the high-lying $e_g$ and low-lying $t_{2g}$ orbitals. As displayed in **Fig. 1b**, the three left electrons of Mn atom would half-fill the $t_{2g}$ orbitals, giving rise to a magnetic moment of 3 $\mu_B$. As expected, our spin-polarized calculations show that P+ and P- phases are spin-polarized and the magnetic moments are mainly distributed on the Mn atoms.

**Table 1.** Magnetic parameters of monolayer MnSTe, P+ phase and P- phase.

|        | $D$ (meV) | $J$ (meV) | $\lambda$ (meV) | $K$ (meV) | $D/J$ | $D^2/|KJ|$ |
|--------|-----------|-----------|-----------------|-----------|-------|------------|
| MnSTe  | 1.875     | 10.228    | -0.034          | 0.401     | 0.183 | 0.857      |
| P+     | 2.181     | 9.369     | -0.238          | 0.393     | 0.233 | 1.292      |
| P-     | 2.520     | 8.086     | -0.320          | 0.415     | 0.312 | 1.892      |

To investigate the magnetic interactions of the P+ and P- phases, we adopt the following spin Hamiltonian:

$$H = -\sum_{<i,j>} \boldsymbol{D}_{ij} \cdot (\boldsymbol{S}_i \times \boldsymbol{S}_j) - J \sum_{<i,j>} (\boldsymbol{S}_i \cdot \boldsymbol{S}_j) - \lambda \sum_{<i,j>} (S_i^z \cdot S_j^z) - K \sum_i (S_i^z)^2 - mB \sum_i S_i^z \quad (1)$$

Here, $\boldsymbol{S}_i$ and $\boldsymbol{S}_j$ are the unit vectors representing local spins of the $i^{th}$ and $j^{th}$ Mn atoms, respectively. $\boldsymbol{D}_{ij}$ is the Dzyaloshinskii-Moriya interaction (DMI), $J$ is the isotropic exchange coupling, $\lambda$ is anisotropic exchange coupling, and $K$ is the single ion anisotropy. The last term is the Zeeman energy. Based on the Moriya's rule[58], because there is a mirror plane passing through the middle of the bond between two neighboring Mn atoms, $\boldsymbol{D}_{ij}$ for the nearest-neighboring Mn atoms is perpendicular to their bond. As the out-of-plane component of $\boldsymbol{D}_{ij}$ is negligible, we only consider its in-plane component. To calculate the DMI parameter, a $4 \times 1$ supercell is constructed, and two different noncollinear magnetic orders are set [i.e., left-hand and right-hand orders, see **Fig. 2a**]. The calculated $\boldsymbol{D}_{ij}$ are listed in **Table 1**. For free-standing MnSTe, $\boldsymbol{D}_{ij}$ is calculated to be 1.875 meV. Such a large value can be attributed to the combined effect of the large electronegativity difference between S and Te atoms and strong spin-orbital coupling (SOC) strength within Te atom. By forming the heterobilayers, due to the interfacing effect, $\boldsymbol{D}_{ij}$ is further enhanced to 2.181 and 2.502 meV for P+ and P- phases, respectively. The significantly difference in $\boldsymbol{D}_{ij}$ for P+ and P- can be attributed to their distinct discrepancy in electric polarization.

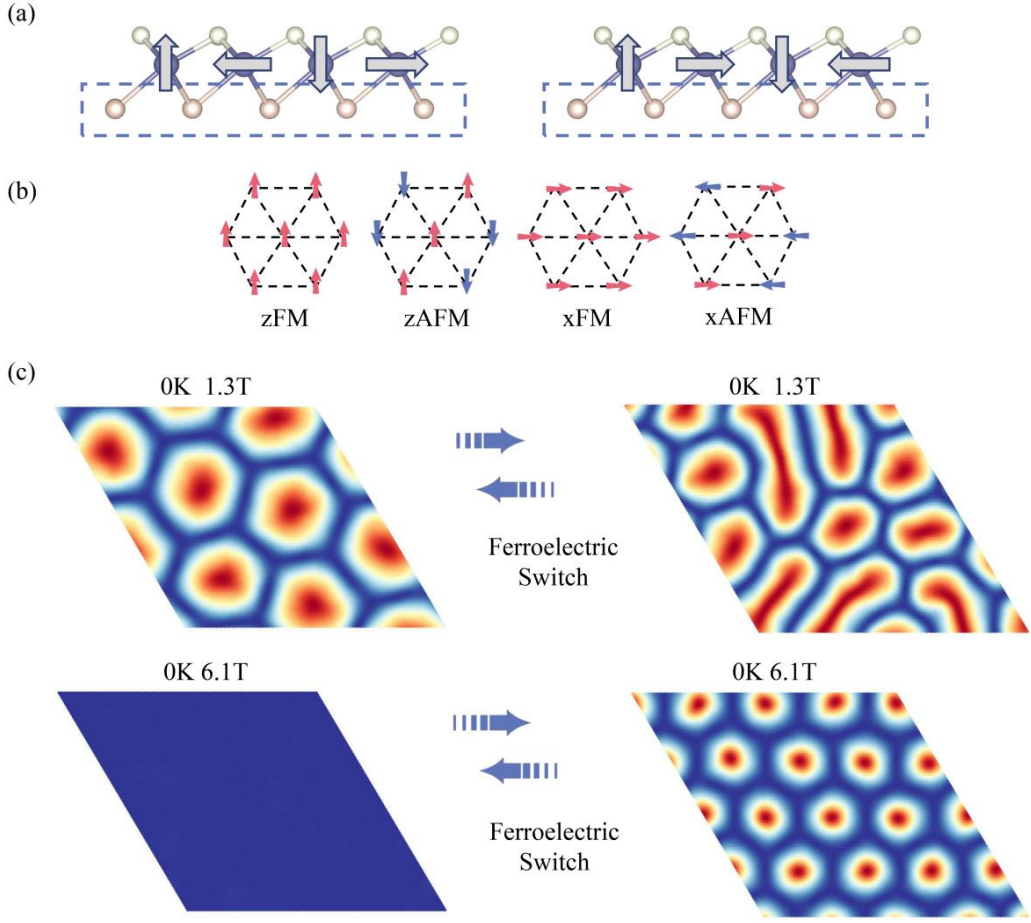

**Fig. 2** (a) Illustrations of right- and left-hand spin-spiral configurations adopted to obtain $D_{ij}$. (b) Illustrations of spin configurations used to estimate $J$, $K$ and $\lambda$. (c) Spin textures for P+ (left panel) and P- (right panel) phases under external magnetic field.

To get $J$, $K$ and $\lambda$, four magnetic configurations illustrated in **Fig. 2b** are considered, and the corresponding results are listed in **Table 1**. The isotropic exchange interaction $J$ is calculated to be 9.369 and 8.086 meV, respectively, for P+ and P- phases, suggesting ferromagnetic (FM) exchange interaction among the magnetic moments. Such FM coupling is related to the super exchange mechanism. As shown in **Fig. 1b**, for the electron in $p$ orbitals of S/Te atom, it would couple anti-parallelly with the electrons in $d$ orbitals, as the low-lying $t_{2g}$ orbitals are fully half-filled. As a result, mediated by $p$ orbitals of S/Te atoms, the coupling among Mn atoms in both P+ and P- phases favor FM coupling. In fact, FM ground state coupling in the heterobilayers can also be deduced from their crystal structures. The Mn-S(Te)-Mn bonding angle is found to be 97.3° (81.8°) and 97.1° (81.6°) for P+ and P- phases, respectively, which are close to 90.0°. According to the Goodenough-Kanamori-Anderson rules[59], FM coupling should dominate the exchange interaction among Mn atoms. Also, similar with the case of DMI, the different *J* values in P+ and P- phases correlate to their opposite electric polarizations.

Different from the cases of DMI and isotropic exchange interaction, as listed in **Table 1**, the anisotropic exchange coupling *λ* is significantly enhanced in the heterobilayers with respect to that of monolayer MnSTe. The negative values of *λ* indicate that the Heisenberg exchange interaction favors the in-plane magnetization for P+ and P- phases. However, as compared with the Heisenberg exchange interaction, the single ion anisotropy *K* is obviously larger. As listed in **Table 1**, the single ion anisotropy exhibits positive values, which suggests our-of-plane magnetization. The competition between them leads to the out-of-plane magnetization orientation to be more favorable for both systems.

By featuring different magnetic parameters, different spin textures are expected in P+ and P- phases, indicating the possibility for realizing ferroelectrically controllable magnetic behaviors. To verify this scenario, parallel tempering Monte Carlo (MC) simulations[56] are performed to explore the spin textures in the heterobilayers, which is based on the first-principles calculations parametrized Hamiltonian in **Eq. 1**. The spin textures for P+ and P- phases in the absence of an external magnetic field are presented in **Fig. S4**, which display a ferromagnetic state with labyrinth domains. Such labyrinth domains can be broken when applying external magnetic field. Remarkably, we find that under an external magnetic field of 1.3 T, the labyrinth pattern is transformed into the intriguing skyrmion lattice state in P+ phase; see **Fig. 2c**. Besides, the radius of skyrmion is ~15 nm. In contrast to P+ phase, although some isolated skyrmions also appear in P- phase under the magnetic field of 1.3 T, the fragmented labyrinth domains are observed simultaneously, as shown in **Fig 2c**. Such mixed phase of skyrmion and fragmented labyrinth domain in P- phase renders the isolated skyrmions in it technologically useless. As P+ and P- phases can be considered as two ferroelectric states of the heterobilayer, the skyrmion lattice phase can be switched on and off by ferroelectricity, thus realizing the strong magnetoelectric effect of ferroelectrically controllable skyrmionics.

Moreover, by applying an external magnetic field of 6.1 T, as shown in **Fig. 2c**, neither fragmented labyrinth domain nor skyrmion lattice is observed in P+ phase, giving rise to a trivial ferromagnetism state. While for P- phase under an external magnetic field of 6.1 T, interestingly, the labyrinth pattern transforms into skyrmion lattice state, and the radius of skyrmion is found to be ~8 nm; see **Fig. 2c**. Apparently, similar to the case of 1.3 T, the skyrmion phase can be switched on and off by reversing the electric polarization, leading to the ferroelectrically controllable skyrmionics. However, the "switch-on" and "switch-off" states are reversed with respected to that of 1.3 T.

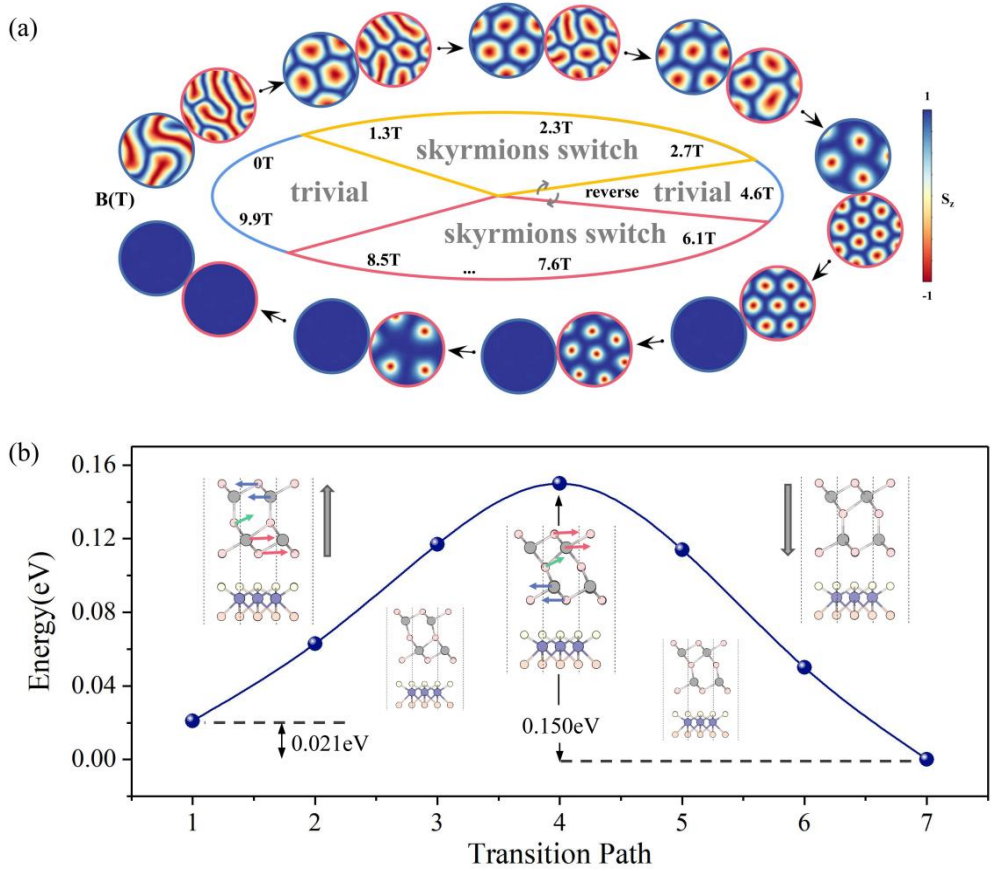

**Fig. 3** (a) Spin textures for P+ (with blue edge) and P- (with red edge) phases under various external magnetic fields. (b) Ferroelectric transition pathway and energy barrier between P+ and P- phases of the heterobilayer.

To explore the robustness of such magnetoelectric effect in terms of external magnetic field, we further investigate their spin textures under various magnetic fields. The corresponding results are presented in **Fig. 3a**. Interestingly, under the external magnetic fields of 1.3 - 3.6 T and 6.1 - 8.5 T, the magnetoelectric effect of ferroelectrically controllable skyrmionics in this heterobilayer can be preserved, suggesting the phenomenon is rather robust. For the cases of 1.3 - 3.6 T, with increasing the external magnetic field, the density of skyrmions in P+ phase is roughly maintained and the size of skyrmion is reduced, while the fragmented labyrinth domains in P- phase become more and more isolated. For the cases of 6.1 - 8.5 T, with increasing the external magnetic field, while the P+ phase preserves a trivial ferromagnetic state, the density of skyrmions in P- phase decreases and the size of skyrmion is roughly maintained. Under the external magnetic field of 3.6 - 6.1 T, as shown in **Fig. 3a**, the skyrmion lattice state is achieved for both P+ and P- phases, excluding the magnetoelectric effect of ferroelectrically controllable skyrmionics in this heterobilayer. And when increasing the magnetic field larger than 8.5 T, both phases favor the trivial ferromagnetic state, also prohibiting the realization of such magnetoelectric effect.

Having established the magnetoelectric effect of ferroelectrically controllable skyrmionics, we then investigate the feasibility of ferroelectricity in this heterobilayer. To this end, we employ the nudged elastic band (NEB) method to study the ferroelectric switching process. The P+ phase is lower in energy than the P- phase by 21 meV per f.u. of $In_2Se_3$. The ferroelectric transition process between P+ and P- phases is displayed in **Fig. 3c**. The transition between P+ phase and paraelectric state is realized by the left-shifting (right-shifting) of upper (lower) In-Se layer and the upper-inclined-shifting of inner Se layer. And the transition between paraelectric state and P- phases is caused by the right-shifting (left-shifting) of upper (lower) In-Se layer and the upper-inclined-shifting of inner Se layer. The energy barrier for the ferroelectric switching from P+ (P-) to P- (P+) phases is estimated to be 129 (150) meV per f.u. of $In_2Se_3$. These values are significantly smaller than those of $VOCl_2$ (180 meV/f.u.)[60], $Sc_2CO_2$ (520 meV/f.u.)[61], vacancy-doped $CrI_3$ (650 meV/f.u.)[62], but lager than those of NaOH (75 meV/f.u.)[63], $CuCrP_2S_6$ (100 meV/f.u.)[64] indicating the feasibility of the ferroelectricity and thus the strong magnetoelectric effect in this heterobilayer.

To get more physics into the magnetoelectric effect in this heterobilayer, we discuss the relationship between topological spin texture and magnetic parameters of *D*, *J*, and *K*. Among these three parameters, *J* and *K* are inclined to force the magnetic moments to be arranged parallelly, while *D* is prone to induce noncolinear alignments of magnetic moments. Based on this fact, we find that $D^2/|KJ|$ can describe the magnetoelectric effect in this heterobilayer. In detail, $D^2/|KJ|$ can be used to qualitatively estimate the required external magnetic field for realizing the skyrmion state. Taking $D^2/|KJ|$ of 1.892 for P- phase as reference, we set $D^2/|KJ|$ to be 1.135, 1.514, 2.270 and 2.649, respectively, corresponding to 0.6, 0.8, 1.2 and 1.4 times 1.892. The corresponding spin textures and phase diagram for them under external magnetic field are shown in **Fig. 4a,b**. For all these cases, the labyrinth domain state can be transformed into skyrmion lattice state and then to the trivial ferromagnetic state by applying external magnetic field. With decreasing the value of $D^2/|KJ|$, the critical magnetic fields required for such transitions decreases. As P+ and P- phases exhibit different $D^2/|KJ|$, by applying an appropriate external magnetic field, the magnetoelectric effect of ferroelectrically controllable skyrmionics can be realized.

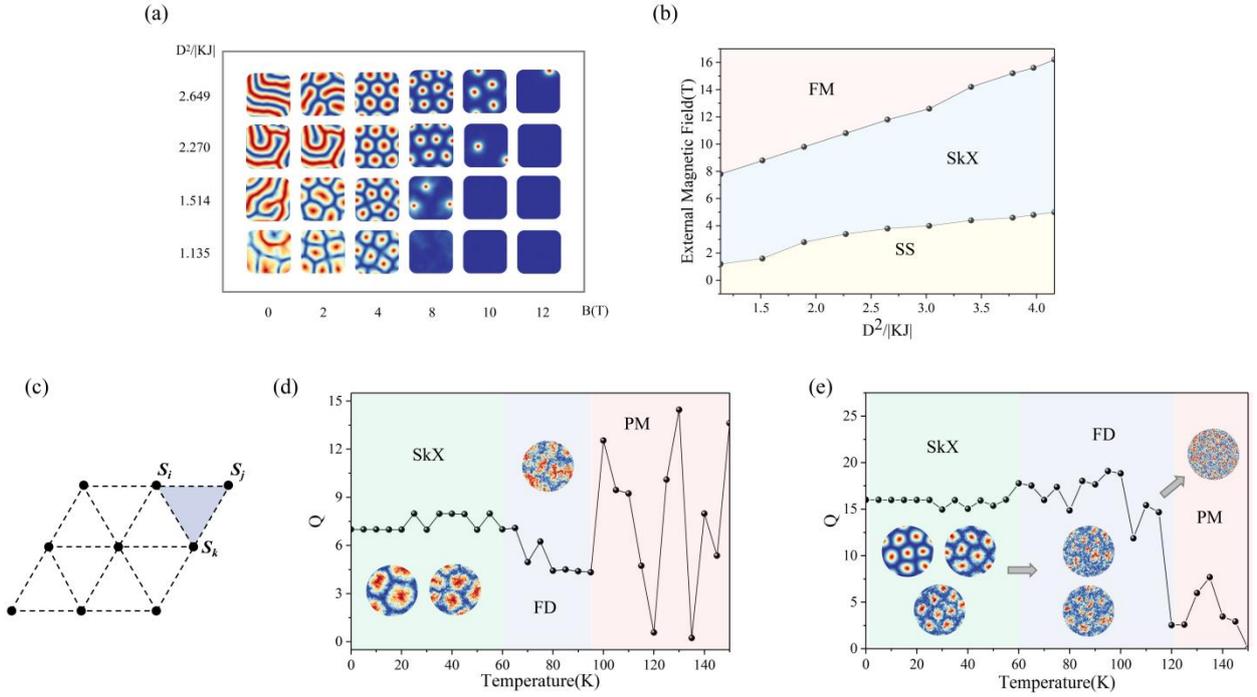

**Fig. 4** (a) Spin textures and (b) phase diagram as functions of external magnetic field and $D^2/|KJ|$. (c) Equilateral triangle lattice employed for calculating skyrmion number Q. The variation of Q as a function of temperature and the corresponding phase diagrams for (d) P+ and (e) P- phases. SkX, FD and PM in (b,d,e) indicate skyrmion, fluctuation disorder and paramagnetic states, respectively.

At last, we take P+ and P- phases under the external magnetic field of 1.3 and 6.1 T, respectively, as examples to discuss the temperature effect on the skyrmion lattice state. To quantitatively characterize such effect, the topological index is employed, which is defined as[65]

$$Q = \frac{1}{4\pi} \int \boldsymbol{m} \cdot \left( \frac{\partial \boldsymbol{m}}{\partial x} \times \frac{\partial \boldsymbol{m}}{\partial y} \right) dxdy$$

Here, m is the normalized magnetization. Q describes the number of unit sphere that the magnetic moments can be wrapped around. Q = ±1 refers one skyrmion. This formula is in continuous integral form, which is not convenient for discrete lattice. According to the method proposed by Berg and Luscher et al.[65], we partition the lattice into many unit right triangles[66], as shown in **Fig. 4c**. All unit triangles are employed to calculate the Q as follow:

$$Q = \frac{1}{4\pi} \sum_l qn$$

$$\tan \frac{qn}{2} = \frac{\boldsymbol{S}_i^n \cdot (\boldsymbol{S}_j^n \times \boldsymbol{S}_k^n)}{1 + \boldsymbol{S}_i^n \cdot \boldsymbol{S}_j^n + \boldsymbol{S}_j^n \cdot \boldsymbol{S}_k^n + \boldsymbol{S}_k^n \cdot \boldsymbol{S}_i^n}$$

Here, $S_i^n$, $S_j^n$, $S_k^n$ are the magnetic moments of the three atoms at the vertices of $n^{th}$ equilateral triangle in the counter clockwise lattice.

The variation of Q as a function of temperature and the corresponding phase diagrams are displayed in **Fig. 4d,e**. For the case of P+, as shown in **Fig. 4d**, the skyrmion lattice state can be preserved with temperature reaching up to 60 K. In the skyrmion lattice state, Q is almost unchanged with increasing temperature, arising from that the edge of skyrmion is becoming vagueness. When further increasing the temperature, the skyrmion lattice state is transformed into the fluctuation disorder (FD) state and finally paramagnetic (PM) state. It should be noted that in FD and PM state, Q oscillates with increasing temperature. This phenomenon can be attributed to the competition between the following two factors. First, since Q is essentially the product of polarity and vorticity, the breaking of skyrmions caused by temperature increases of the number of little vortices in the spin textures, leading to the increase of Q. Second, the temperature-induced vanishing of vortex would decrease Q. The roughly similar scenario is shared by the case of P-; see **Fig. 4e**.

**Conclusion**

In conclusion, using first-principles calculations and Monte-Carlo simulations, we report the discovery of a multiferroic heterobilayer with strong magnetoelectric effect. We reveal that with switching the electric polarization in ferroelectric α-$In_2Se_3$, the creation and annihilation of skyrmion lattice state can be realized, giving rise to the exotic ferroelectrically controllable skyrmionics. Such phenomenon is revealed to be closely related to the physical quantity of $D^2/|KJ|$, and this quantity can describe the required conditions for such physics. In addition, the evolution of Q with temperature in the heterobilayers and the corresponding phase diagram are investigated. These results shed new light on the research of 2D magnetoelectric multiferroics and skyrmionics.

**Acknowledgement**

This work is supported by the National Natural Science Foundation of China (Nos. 11804190 and 12074217), Shandong Provincial Natural Science Foundation (Nos. ZR2019QA011 and ZR2019MEM013), Shandong Provincial Key Research and Development Program (Major Scientific and Technological Innovation Project) (No. 2019JZZY010302), Shandong Provincial Science Foundation for Excellent Young Scholars (No. ZR2020YQ04), and Qilu Young Scholar Program of Shandong University.

**References**

1. Gong, C.; Kim, E. M.; Wang, Y.; Lee, G.; Zhang, X., Multiferroicity in atomic van der Waals heterostructures. *Nat. Commun.* **2019,** *10* (1), 2657.
2. Feng, Y.; Peng, R.; Dai, Y.; Huang, B.; Duan, L.; Ma, Y., Antiferromagnetic ferroelastic multiferroics in single-layer VOX (X = Cl, Br) predicted from first-principles. *Appl. Phys. Lett.* **2021,** *119* (17), 173103.
3. Luo, W.; Xu, K.; Xiang, H., Two-dimensional hyperferroelectric metals: A different route to ferromagnetic- ferroelectric multiferroics. *Phys. Rev. B* **2017,** *96* (23), 235415.
4. Yang, L.; Wu, M.; Yao, K., Transition-metal-doped group-IV monochalcogenides: a combination of two-dimensional triferroics and diluted magnetic semiconductors. *Nanotechnol.* **2018,** *29* (21), 215703.
5. Liang, Y.; Guo, R.; Shen, S.; Huang, B.; Dai, Y.; Ma, Y., Out-of-plane ferroelectricity and multiferroicity in elemental bilayer phosphorene, arsenene, and antimonene. *Appl. Phys. Lett.* **2021,** *118* (1), 012905.
6. Eerenstein, W.; Mathur, N.; Scott, J., Multiferroic and magnetoelectric materials. *Nature* **2006,** *442* (7104), 759-765.
7. Cheong, S.; Mostovoy, M., Multiferroics: a magnetic twist for ferroelectricity. *Nat. Mater.* **2007,** *6*, 13-20.
8. Spaldin, N.; Ramesh, R., Advances in magnetoelectric multiferroics. *Nat. Mater.* **2019,** *18*, 203-212.
9. Xu, M.; Huang, C.; Li, Y.; Liu, S.; Zhong, X.; Jena, P.; Kan, E.; Wang, Y., Electrical control of magnetic phase transition in a type-I multiferroic double-metal trihalide monolayer. *Phys. Rev. Lett.* **2020,** *124* (6), 067602.
10. Giraldo, M.; Meier, Q.; Bortis, A.; Nowak, D.; Spaldin, N.; Fiebig, M.; Weber, M.; Lottermoser, T., Magnetoelectric coupling of domains, domain walls and vortices in a multiferroic with independent magnetic and electric order. *Nat. Commun.* **2021,** *12* (1), 3093.
11. Zhang, J.; Lin, L.; Zhang, Y.; Wu, M.; Yakobson, B.; Dong, S., Type-II multiferroic $Hf_2VC_2F_2$ MXene monolayer with high transition temperature. *J Am. Chem. Soc.* **2018,** *140* (30), 9768-9773.
12. Shen, S.; Liu, C.; Ma, Y.; Huang, B.; Dai, Y., Robust two-dimensional ferroelectricity in single-layer γ-SbP and γ-SbAs. *Nanoscale* **2019,** *11* (24), 11864-11871.
13. Feng, D.; Zhu, Z.; Chen, X.; Qi, J., Electric-polarization-driven magnetic phase transition in a ferroelectric- ferromagnetic heterostructure. *Appl. Phys. Lett.* **2021,** *118* (6), 062903.
14. Yang, B.; Shao, B.; Wang, J.; Li, Y.; Yam, C.; Zhang, S.; Huang, B., Realization of semiconducting layered multiferroic heterojunctions via asymmetrical magnetoelectric coupling. *Phys. Rev. B* **2021,** *103* (20), L201405.
15. Sun, W.; Wang, W.; Chen, D.; Cheng, Z.; Wang, Y., Valence mediated tunable magnetism and electronic properties by ferroelectric polarization switching in 2D $FeI_2/In_2Se_3$ van der Waals heterostructures. *Nanoscale* **2019,** *11* (20), 9931-9936.
16. Xue, F.; Wang, Z.; Hou, Y.; Gu, L.; Wu, R., Control of magnetic properties of $MnBi_2Te_4$ using a van der Waals ferroelectric $III_2$-$VI_3$ film and biaxial strain. *Phys. Rev. B* **2020,** *101* (18), 184426.
17. Xu, S.; Jia, F.; Zhao, G.; Wu, W.; Ren, W., Two-dimensional ferroelectric ferromagnetic half semiconductor in VOF monolayer. *J. Mater. Chem. C* **2021,** *9*, 9130-9136.
18. Shang, J.; Li, C.; Tang, X.; Du, A.; Liao, T.; Gu, Y.; Ma, Y.; Kou, L.; Chen, C., Multiferroic decorated $Fe_2O_3$ monolayer predicted from first principles. *Nanoscale* **2020,** *12* (27), 14847-14852.

19. Huang, C.; Du, Y.; Wu, H.; Xiang, H.; Kaiming Deng; Kan, E., Prediction of intrinsic ferromagnetic ferroelectricity in a transition-metal halide monolayer. *Phys. Rev. Lett.* **2018,** *120* (6), 147601.
20. Sun, W.; Wang, W.; Li, H.; Zhang, G.; Chen, D.; Wang, J.; Cheng, Z., Controlling bimerons as skyrmion analogues by ferroelectric polarization in 2D van der Waals multiferroic heterostructures. *Nat. Commun.* **2020,** *11* (1), 5930.
21. Cui, Q.; Zhu, Y.; Jiang, J.; Liang, J.; Yu, D.; Cui, P.; Yang, H., Ferroelectrically controlled topological magnetic phase in a Janus-magnet-based multiferroic heterostructure. *Phys. Rev. Research* **2021,** *3* (4), 043011.
22. Li, C.; Yao, X.; Chen, G., Writing and deleting skyrmions with electric fields in a multiferroic heterostructure. *Phys. Rev. Research* **2021,** *3* (1), L012026.
23. Khomskii, D., Classifying multiferroics: Mechanisms and effects. *Physics* **2009,** *2* (20).
24. Kang, W.; Huang, Y.; Zheng, C.; Lv, W.; Lei, N.; Zhang, Y.; Zhang, X.; Zhou, Y.; Zhao, W., Voltage controlled magnetic skyrmion motion for racetrack memory. *Sci. Rep.* **2016,** *6*, 23164.
25. Parking, S.; Hayashi, M.; Thomas, L., Magnetic domain-wall racetrack memory. *Science* **2008,** *320* (5873), 190-194.
26. Fert, A.; Cros, V.; Sampaio, J., Skyrmions on the track. *Nat. Nanotechnol.* **2013,** *8* (3), 152-156.
27. Mühlbauer, S.; Binz, B.; Jonietz, F.; Pfleiderer, C.; Rosch, A.; Neubauer, A.; Georgii, R.; Böni, P., Skyrmion lattice in a chiral magnet. *Science* **2009,** *323* (5916), 915-919.
28. Tokura, Y.; Kanazawa, N., Magnetic skyrmion materials. *Chem. Rev.* **2021,** *121* (5), 2857-2897.
29. Jiang, W.; Chen, G.; Liu, K.; Zang, J.; Velthuis, S.; Hoffmann, A., Skyrmions in magnetic multilayers. *Phys. Rep.* **2017,** *704*, 1-49.
30. Yu, X.; Onose, Y.; Kimoto, K.; Zhang, W.; Ishiwata, S.; Matsui, Y.; Tokura, Y., Near room-temperature formation of a skyrmion crystal in thin-films of the helimagnet FeGe. *Nat. Mater.* **2011,** *10* (2), 106-109.
31. Huang, B.; Clark, G.; Moratalla, E.; Klein, D.; Cheng, R.; Seyler, K.; Zhong, D.; Schmidgall, E.; McGuire, M.; Cobden, D.; Yao, W.; Xiao, D.; Herrero, P.; Xu, X., Layer-dependent ferromagnetism in a van der Waals crystal down to the monolayer limit. *Nature* **2017,** *546* (7657), 270-273.
32. Gong, C.; Li, L.; Li, Z.; Ji, H.; Stern, A.; Xia, Y.; Cao, T.; Bao, W.; Wang, C.; Wang, Y.; Qiu, Z.; Cava, R.; Louie, S.; Xia, J.; Zhang, X., Discovery of intrinsic ferromagnetism in two-dimensional van der Waals crystals. *Nature* **2017,** *546* (7657), 265-269.
33. Bonilla, M.; Kolekar, S.; Ma, Y.; Diaz, H.; Kalappattil, V.; Das, R.; Eggers, T.; Gutierrez, H.; Phan, M.; Batzill, M., Strong room-temperature ferromagnetism in $VSe_2$ monolayers on van der Waals substrates. *Nat. Nanotechnol.* **2018,** *13* (4), 289-293.
34. Ma, Y.; Dai, Y.; Guo, M.; Niu, C.; Zhu, Y.; Huang, B., Evidence of the existence of magnetism in pristine $VX_2$ Monolayers (X = S, Se) and their strain-induced tunable magnetic properties. *ACS Nano* **2012,** *6* (2), 1695-1701.
35. Yuan, J.; Yang, Y.; Cai, Y.; Wu, Y.; Chen, Y.; Yan, X.; Shen, L., Intrinsic skyrmions in monolayer Janus magnets. *Phys. Rev. B* **2020,** *101* (9), 094420.
36. Liang, J.; Wang, W.; Du, H.; Hallal, A.; Garcia, K.; Chshiev, M.; Fert, A.; Yang, H., Very large Dzyaloshinskii- Moriya interaction in two-dimensional Janus manganese dichalcogenides and its application to realize skyrmion states. *Phys. Rev. B* **2020,** *101* (18), 184401.
37. Cui, Q.; Liang, J.; Shao, Z.; Cui, P.; Yang, H., Strain-tunable ferromagnetism and chiral spin textures in two-dimensional Janus chromium dichalcogenides. *Phys. Rev. B* **2020,** *102* (9), 094425.
38. Zhang, Y.; Xu, C.; Chen, P.; Nahas, Y.; Prokhorenko, S.; Bellaiche, L., Emergence of skyrmionium in a two- dimensional $CrGe(Se,Te)_3$ Janus monolayer. *Phys. Rev. B* **2020,** *102* (24), 241107.

39. Xu, C.; Feng, J.; Prokhorenko, S.; Nahas, Y.; Xiang, H.; Bellaiche, L., Topological spin texture in Janus monolayers of the chromium trihalides Cr(I, X)$_3$. *Phys. Rev. B* **2020,** *101* (6), 060404.

40. Liang, J.; Cui, Q.; Yang, H., Electrically switchable Rashba-type Dzyaloshinskii-Moriya interaction and skyrmion in two-dimensional magnetoelectric multiferroics. *Phys. Rev. B* **2020,** *102* (22), 220409.

41. Xu, C.; Chen, P.; Tan, H.; Yang, Y.; Xiang, H.; Bellaiche, L., Electric-field switching of magnetic topological charge in type-I multiferroics. *Phys. Rev. Lett.* **2020,** *125* (3), 037203.

42. Amoroso, D.; Picozzi, S., Spontaneous skyrmionic lattice from anisotropic symmetric exchange in a Ni-halide monolayer. *Nat. Commun.* **2020,** *11* (1), 5784.

43. Jiang, J.; Liu, X.; Li, R.; Mi, W., Topological spin textures in a two-dimensional MnBi$_2$(Se, Te)$_4$ Janus material. *Appl. Phys. Lett.* **2021,** *119* (7), 072401.

44. Zhang, X.; Zhou, Y.; Mee Song, K.; Park, T.; Xia, J.; Ezawa, M.; Liu, X.; Zhao, W.; Zhao, G.; Woo, S., Skyrmion-electronics: writing, deleting, reading and processing magnetic skyrmions toward spintronic applications. *J. Phys.: Condens. Matter* **2020,** *32* (14), 143001.

45. Rosch, A., Electric control of skyrmions. *Nat. Nanotechnol.* **2016,** *12* (2), 103-104.

46. Schott, M.; Bernand-Mantel, A.; Ranno, L.; Pizzini, S.; Vogel, J.; Béa, H.; Baraduc, C.; Auffret, S.; Gaudin, G.; Givord, D., The skyrmion switch: Turning magnetic skyrmion bubbles on and off with an electric field. *Nano Lett.* **2017,** *17* (5), 3006-3012.

47. Wang, J.; Strungaru, M.; Ruta, S.; Meo, A.; Zhou, Y.; Deák, A.; Szunyogh, L.; Gavriloaea, P.; Moreno, R.; Chubykalo-Fesenko, O.; Wu, J.; Xu, Y.; Evans, R.; Chantrell, R., Spontaneous creation and annihilation dynamics of magnetic skyrmions at elevated temperature. *Phys. Rev. B* **2021,** *104* (5), 054420.

48. Sampaio, J.; Cros, V.; Rohart, S.; Thiaville, A.; Fert, A., Nucleation, stability and current-induced motion of isolated magnetic skyrmions in nanostructures. *Nat. Nanotechnol.* **2013,** *8* (11), 839-844.

49. Kresse, G.; Furthmuller, J., Efficient iterative schemes for ab initio total-energy calculations using a plane-wave basis set. *Phys. Rev. B* **1996,** *54* (16), 11169-11186.

50. Blochl, P., Projector augmented-wave method. *Phys. Rev. B Condens Matter* **1994,** *50* (24), 17953-17979.

51. John P.; Kieron B.; Ernzerhof, M., Generalized gradient approximation made simple. *Phys. Rev. Lett.* **1996,** *77* (18), 3865-3868.

52. Anisimov, V.; Aryasetiawan, F.; Lichtenstein, A., First-principles calculations of the electronic structure and spectra of strongly correlated systems: the LDA+ U method. *J. Phys.: Condens. Matter* **1997,** *9*, 767-808.

53. Monkhorst, H.; Pack, J., Special points for Brillouin-zone integrations. *Phys. Rev. B* **1976,** *13* (12), 5188-5192.

54. Grimme, S.; Antony, J.; Ehrlich, S.; Krieg, H., A consistent and accurate ab initio parametrization of density functional dispersion correction (DFT-D) for the 94 elements H-Pu. *J Chem. Phys.* **2010,** *132* (15), 154104.

55. Gregory M.; Gregory K., Reversible work transition state theory: application to dissociative adsorption of hydrogen. *Surface Science* **1995,** *324*, 305-337.

56. Miyatake, Y.; Yamamoto, M.; Kim, J.; Toyonaga, M.; Nagai, O., On the implementation of the 'heat bath' algorithms for Monte Carlo simulations of classical Heisenberg spin systems. *J. Phys. C: Solid State Phys.* **1986,** *19*, 2539-2546.

57. Ding, W.; Zhu, J.; Wang, Z.; Gao, Y.; Xiao, D.; Gu, Y.; Zhang, Z.; Zhu, W., Prediction of intrinsic two-dimensional ferroelectrics in In$_2$Se$_3$ and other III$_2$-VI$_3$ van der Waals materials. *Nat. Commun.* **2017,** *8* (1), 14956.

58. Moriya, T., Anisotropic superexchange interaction and weak ferromagnetism. *Phys. Rev.* **1960,** *120* (1), 91-98.

59. Goodenough, J., Theory of the role of covalence in the perovskite-type manganites [La, M(II)]MnO$_3$. *Phys. Rev.* **1955,** *100* (2), 564-573.

60. Ai, H.; Song, X.; Qi, S.; Li, W.; Zhao, M., Intrinsic multiferroicity in two-dimensional VOCl$_2$ monolayers. *Nanoscale* **2019,** *11* (3), 1103-1110.

61. Chandrasekaran, A.; Mishra, A.; Singh, A., Ferroelectricity, antiferroelectricity, and ultrathin 2D electron/hole gas in multifunctional monolayer MXene. *Nano Lett.* **2017,** *17* (5), 3290-3296.

62. Zhao, Y.; Lin, L.; Zhou, Q.; Li, Y.; Yuan, S.; Chen, Q.; Dong, S.; Wang, J., Surface Vacancy-Induced Switchable Electric Polarization and Enhanced Ferromagnetism in Monolayer Metal Trihalides. *Nano Lett.* **2018,** *18* (5), 2943-2949.

63. Ren, Y.; Dong, S.; Wu, M., Unusual ferroelectricity of trans-unitcell ion-displacement and multiferroic soliton in sodium and potassium hydroxides. *ACS Appl. Mater. Inter.* **2018,** *10* (41), 35361-35366.

64. Qi, J.; Wang, H.; Chen, X.; Qian, X., Two-dimensional multiferroic semiconductors with coexisting ferroelectricity and ferromagnetism. *Appl. Phys. Lett.* **2018,** *113* (4), 043102.

65. Berg, B.; Luscher, M., Definition and statistical distributions of a topological number in the lattice O(3) σ-model. *Nuclear Phys. B* **1981,** *190*, 412-424.

66. Rosales, H.; Cabra, D.; Pujol, P., Three-sublattice skyrmion crystal in the antiferromagnetic triangular lattice. *Phys. Rev. B* **2015,** *92* (21), 214439.